\documentclass{PoS}
\usepackage{multirow}
\usepackage{amsmath}
\usepackage{setspace}

\def\NssN{\langle N|\bar{s} s|N \rangle}
\def\fTs{f_{T_s}}
\newcommand{\be}{\begin{equation}}
\newcommand{\ee}{\end{equation}}
\newcommand{\bmin}{\begin{minipage}{0.6\textwidth}}
\newcommand{\emin}{\end{minipage}}

\title{Nucleon mass and strange content from (2+1)-flavor Domain Wall Fermion}

\ShortTitle{Nucleon mass and strange content from (2+1)-flavor Domain Wall Fermion}

\author{\speaker{Chulwoo Jung}%
         \thanks{
We thank RBC and UKQCD Collaborations, especially Yasumichi Aoki, Tom Blum, Chris Dawson, Taku Izubuchi, Meifeng Lin, Shigemi Ohta, Shoichi Sasaki and Takeshi Yamazaki.
}\\
        Physics Department, Brookhaven National Laboratory, Upton, NY 11973, U.S.A.\\
        E-mail: \email{chulwoo@bnl.gov}}
\author{for RBC and UKQCD Collaborations}

\abstract{
We report on our ongoing study of nucleon properties, the nucleon mass and the strange quark content ($\NssN$) in particular.
These calculations are done on (2+1)-flavor Domain Wall Fermion (DWF) ensembles with Iwasaki gauge action with 2 different lattice spacings~(a $\sim$ 0.08, 0.11fm)  and DWF with Dislocation Suppressing Determinant Ratio~(DSDR)~(a $\sim$ 0.14fm), generated by RBC/UKQCD collaborations.
}

\FullConference{The 30th International Symposium on Lattice Field Theory\\
		 June 24 - 29,  2012\\
		 Cairns, Australia}

\begin{document}

\section{Introduction}

Strange content of Nucleon ($\NssN$) and related quantities 
\be
f_{T_s} = \frac{m_s \langle N| \bar{s}s |N \rangle}{m_N}
= \frac{dm_N}{dm_s} \times \frac{m_s}{m_N}, \quad\
\sigma_s = m_s \langle N| \bar{s}s |N \rangle
\ee
has been drawing much attention in the last few years, due to its implications to the dark matter candidate search \cite{Ellis:2008hf}. 
There have been a number of studies of $\NssN$ on the lattice using different approaches such as direct measurement of $\NssN$ matrix element with various noise reduction techniques for disconnected diagrams~\cite{Dinter:2011ns,Gong:2012nw,Engelhardt:2012gd}, indirect measurements from fitting to ChPT formulas for the nucleon mass~\cite{Shanahan:2012wh,Durr:2011mp,Horsley:2011wr}, as well as numerical derivative via Feynman-Hellman theorem \cite{Freeman:2012ry,Oksuzian:2012rzb}. 
A survey of the recent results from lattice shows there is noticeable disagreements between studies with relatively smaller errors
while other studies with multiple lattice spacings often have $\sim$ 100\% relative errors.
This suggests the systematic errors in $\NssN$ may not be well understood.

Here we present  the result of $\NssN$ measurements from numerical derivative via reweighting~\cite{Hasenfratz:2008fg,Liu:2012gm}. Strange quark reweighting has been successfully used to eliminate the systematic error from the discrepancy between dynamical strange quark mass $(m_s)$ and  the physical strange quark mass, and has shown to be usable in shifting $m_s$ by $\sim$ 20\% for the lattice volumes studied~\cite{Aoki:2010dy,Arthur:2012yc}.
\section{Measurement Details}

Table~\ref{table:all} shows the details of the ensembles  and measurements used in this study.
We used (2+1) dynamical flavor DWF with Iwasaki gauge action generated by RBC/UKQCD Collaborations with $a \sim 0.08$, 0.11fm~\cite{Aoki:2010dy}, 
which we will refer to as DWF+I 0.08fm and 0.11fm ensemble respectively. We also measured nucleon mass and $\NssN$ on the DWF ensemble generated with Dislocation Suppressing Determinant Ratio(DSDR~\cite{Renfrew:2009wu}) in addition to the Iwasaki gauge action (DWF+ID). 
Quantities necessary for continuum extrapolation of $\NssN$ such as lattice spacing  and mass renormalization constant were calculated by a combined fit of DWF+I and DWF+ID ensemble described in \cite{Arthur:2012yc}.

For DWF+I 0.11fm ensemble, it was observed in~\cite{Aoki:2010dy}
that the combination of Coulomb gauge-fixed box source of size 16 and the point sink gives a good overlap with the ground state. 
We extended the measurement to use 8 different source positions per configuration, given by
\be
\begin{array}{ccc}
{(x,y,z,t)} = & (0-16,0-16,0-16, n \times 8)\mbox{ for n=even,} &\\ 
& (12-28,12-28,12-28,n \times 8)\mbox{ for n=odd}, & n = 0 \cdots 7
\end{array}
\ee
Where periodicity is implied, so x=12-28 means  x>=12 or x<=4.
 These measurements were binned per configuration,  to get the most of the signal while avoiding the possibility of underestimating the autocorrelation. EigCG\cite{Stathopoulos:2007zi} was employed to decrease amount of computation needed for $8 \times 12= 96$ inversions per configuration, resulting in up to a factor of 3 decrease in the iteration number for the lightest mass $m_l=0.005$.

For DWF+I 0.08fm ensemble, we used propagators generated by LHPC with a gaussian source for their study of nucleon matrix elements\cite{Syritsyn:2009mx}.
As it is shown in Figure~\ref{fig:32}, the nucleon propagator to the point sink exhibits a very slow convergence to the asymptotic value for the nucleon mass and  this proved to be particularly problematic when combined with strange quark mass reweighting.
To circumvent this, propagators with unitary sink, which give consistent values and similar errors for unreweighted nucleon masses with point sinks but with smaller $t_{min}$, were used.

Figure~\ref{fig:DSDR} shows the effective mass and reweighted mass in DWF+ID $(a^{-1}\sim 0.14$fm) ensemble. 
It is worthwhile to note
that effective masses from both point and unitary sink shows a very good agreement, in contrast to DWF+I 0.08fm.

\begin{figure}[hbt]
\vspace{-0.05\textwidth}
\begin{center}
\includegraphics[angle=0,width=0.9\textwidth]{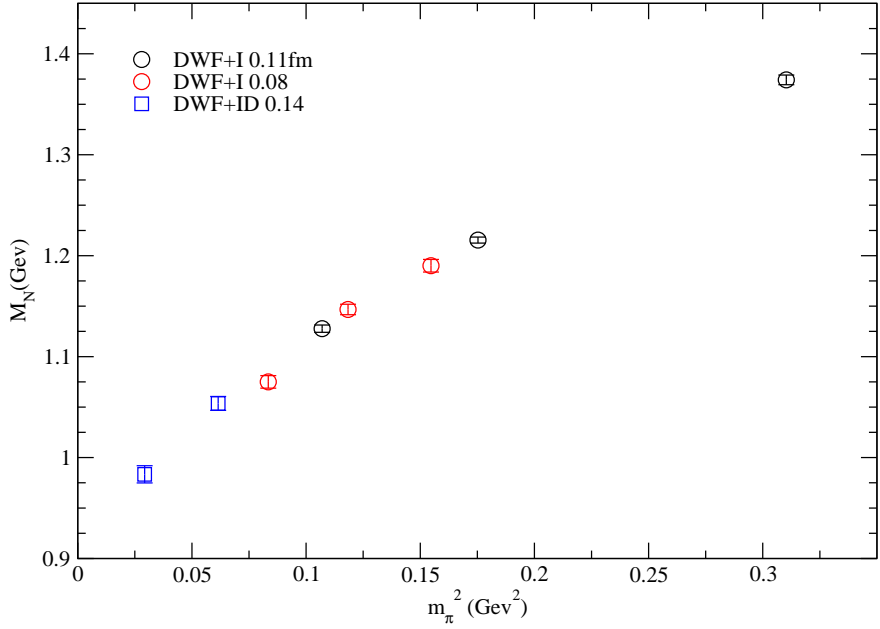}
\end{center}
\vspace{-0.1\textwidth}
\caption{Nucleon masses of the ensembles used in this analysis in physical units.}
\vspace{-0.05\textwidth}
\label{fig:mN}
\end{figure}

Figure~\ref{fig:mN} shows the nucleon mass as a function of pseudoscalar masses squared in physical units. While the lack of measurements with other quantities such as the masses of other octet baryons does not allow us to fit the data with ChPT-motivated fitting forms with any confidence, the smallness of differences in masses  between different ensembles suggests $a^2$ dependence of the nucleon mass is relatively small.

\begin{table}[hbt]
\begin{center}
\begin{tabular}{c|c|c|c|c|c}
\hline
$a m_l$ & $a m_\pi$ &  MD units & measurements  & $t_{min} - t_{max}$ & $a m_N$\\
\hline
\hline
\multicolumn{6}{c}{
DWF+I 0.11fm: $V = 24^3\times 64\times 16,a^{-1} = 1.75(3)$ Gev, $a m_{res} \sim $ 0.003,
}\\
\multicolumn{6}{c}{
$a m'_s = 0.032,0.033, \cdots \underline{0.04}, 0.041, \cdots 0.047$}\\
\hline

0.005   & 0.1891(10) & 1420,1460$\cdots$8980& 1520&  &  0.652(2)\\
0.010   & 0.2420(7) & 1460,1500$\cdots$8540 & 1424& 5-12 & 0.703(2) \\
0.020   & 0.322(1) & 1900,1920$\cdots$3600 & 680   &    & 0.794(3)\\
\hline
\hline

\multicolumn{6}{c}{
DWF+I 0.08fm: $V = 32^3\times 64 \times 16, a^{-1} = 2.31(4) $Gev, $ a m_{res} \sim $ 0.00067,
}\\
\multicolumn{6}{c}{
$a m'_s = 0.0265,0.0270 \cdots \underline{0.03}$.}\\
\hline
0.004   & 0.1267(4) & 590,600$\cdots$6600   & (1996)  &   &    0.474(3)\\
0.006   & 0.1509(3) & 544,552$\cdots$7600   & 3528    &7-15 & 0.503(2) \\
0.008   & 0.1725(5) & 590,600$\cdots$6600   & 2064    &     & 0.524(2)\\
\hline
\hline

\multicolumn{6}{c}{
DWF+ID$ : V= 32^3\times 64 \times 32, a^{-1} = 1.37(1) $Gev, $ a m_{res} \sim $0.0018,
}\\
\multicolumn{6}{c}{
$a m'_s = \underline{0.045},0.0455 \cdots 0.05$.}\\

\hline
0.01    & 0.1250(2) & 500,508$\cdots$2396& (904)   & \multirow{2}{*}{5-11} & 0.718(6) \\
0.042   & 0.1810(2) & 608,616$\cdots$1920 & 1320  & & 0.769(5)\\
\end{tabular}
\end{center}
\caption{Measurement details of ensembles used in the analysis. Parentheses in the number of measurements denotes the measurements are not complete. Underlines for $a m'_s$ denotes the simulated strange quark mass.}
\label{table:all}
\end{table}

\vspace{-0.02\textwidth}
\section{$\NssN$ via reweighting}
\label{section:NssN}
Nucleon strange content is calculated by applying Feynman-Hellman theorem
\be
\NssN = \frac{dM_N(m'_s)}{Z_m dm'_s}
\ee
to nucleon masses for reweighted sea strange mass $M_N(m'_s)$, which is 
calculated by fitting the effective masses calculated from reweighted nucleon propagator $P(t,m'_s)$ to a constant for $t = t_{min}\cdots t_{max}$ for each ensemble.

$P(t,m'_s)$ is calculated by taking the weighted average of the nucleon propagator $P(t,m_s)$ with the weight $W(m'_s,m_s)$, 
which is calculated by multiplying reweighting factors for smaller steps in mass
\begin{gather}
\left< P  (t,m_s') \right>= \frac{\Sigma_i P([U_i],t,m_s) W_i(m'_s,m_s)}{\Sigma_i W_i(m'_s,m_s)},\quad\
W_i(m'_s,m_s) = w_i(m'_s=m_n,m_{n-1}) \cdots w_i(m_2,m_1=m_s). \nonumber 
\end{gather}
Using smaller steps decreases the stochastic noise in the reweighting factor, and also avoids possible difficulties with non-gaussian distribution of the estimated reweighting factor.
Reweighting factors $w_i$ are calculated by estimating the ratio of strange quark determinants with 
Gaussian random vector $\xi$.
\begin{gather}
\Omega_i(m_n',m_n) = D'^{-1}_i D_i D^\dagger_i(D'^{\dagger}_i )^{-1}, 
D_i = D([U_i],m_l,m_n), D'_i = ([U_i],m_l,m_n'), \nonumber \\
w_i(m_n',m_n) = \mbox{det}(\Omega_i(m_n',m_n))^{-1/2}
= \frac{\int d\xi d\xi^\dagger e^{-\xi\dagger
\sqrt{\Omega_i(m_n',m_n)} \xi} } {\int d\xi d\xi^\dagger e^{-\xi\dagger \xi}}  = 
\left<e^{-\xi^\dagger(\sqrt{\Omega_i(m_n',m_n)} -1)\xi} \right>_{\xi}.
\label{eq:rat_quo}
\end{gather}
$\sqrt{\Omega_i(m_n',m_n)}$ is calculated using rational approximation, similar to what is used for the simulation of strange quark in~\cite{Aoki:2010dy,Arthur:2012yc}. 
Using rational approximation makes Eq.~(\ref{eq:rat_quo}) unbiased and allows us to use relatively noisy estimates of the reweighting factors for each mass steps.

$\NssN$ for each ensemble is calculated by correlated fits to 
\be
M_N(m_s',m_l,a) = c'_0 + \NssN(m_l,a) m_s', \\
\ee
then the continuum and chiral extrapolation is taken by fitting to 
\be
\NssN(m_l,a) = c_0 + c_1 m_l( + c_2 a^2).
\label{eq:NssN}
\ee
We used fits both with and without $c_2 a^2$ in Eq.~(\ref{eq:NssN}), to estimate systematic error from the  continuum extrapolation.
\begin{figure}[hbt]
\hspace{-0.05\textwidth}
\bmin
\includegraphics[angle=0,width=1.0\textwidth]{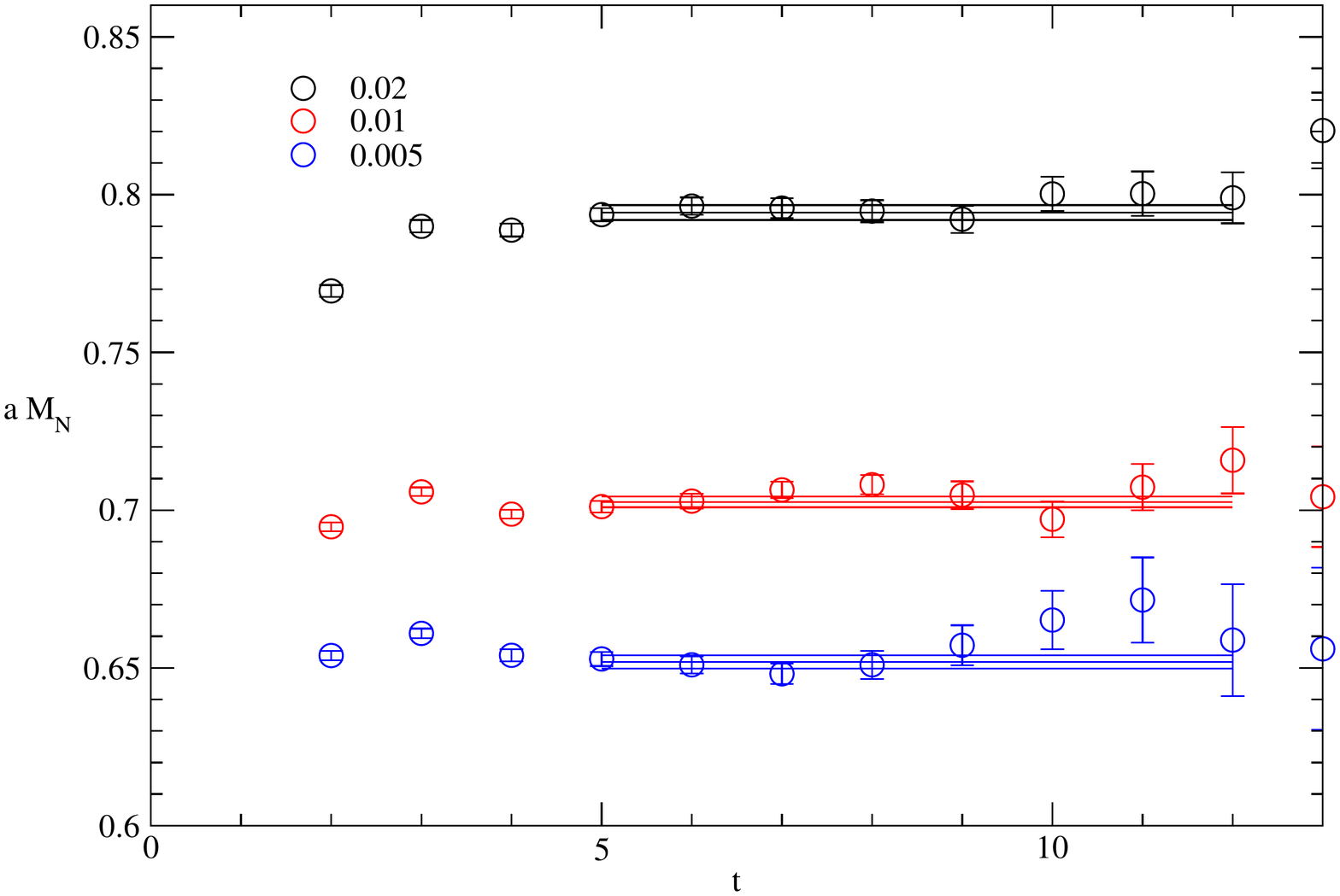}
\emin
\hspace{-0.05\textwidth}
\bmin
\includegraphics[angle=0,width=1.0\textwidth]{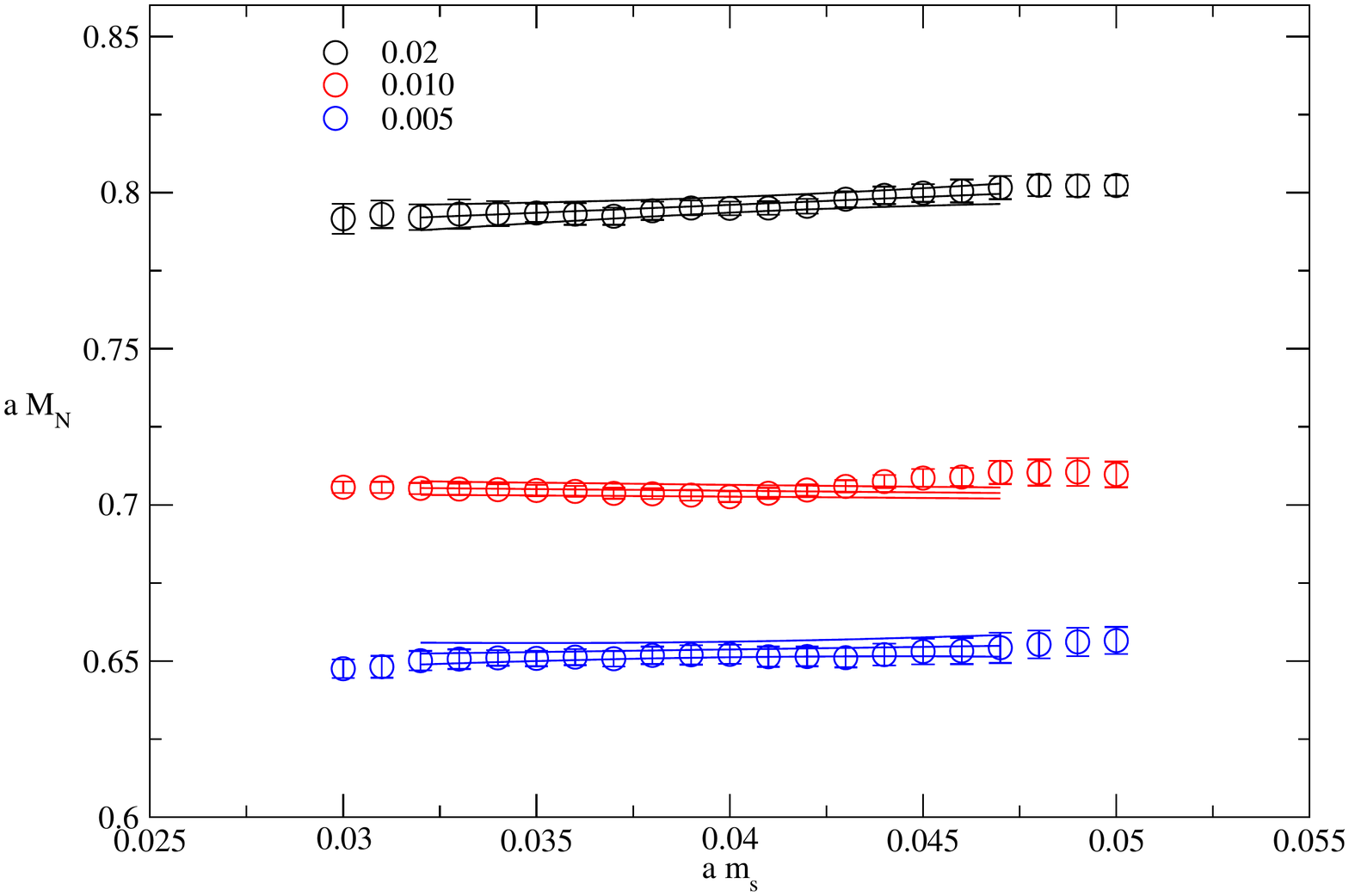}
\emin
\vspace{-0.05\textwidth}
\caption{ Nucleon effective mass and reweighted mass for DWF+I 0.11fm ensembles}
\label{fig:24}
\end{figure}

\begin{figure}[hbt]
\vspace{-0.05\textwidth}
\hspace{-0.05\textwidth}
\bmin
\includegraphics[angle=0,width=1.0\textwidth]{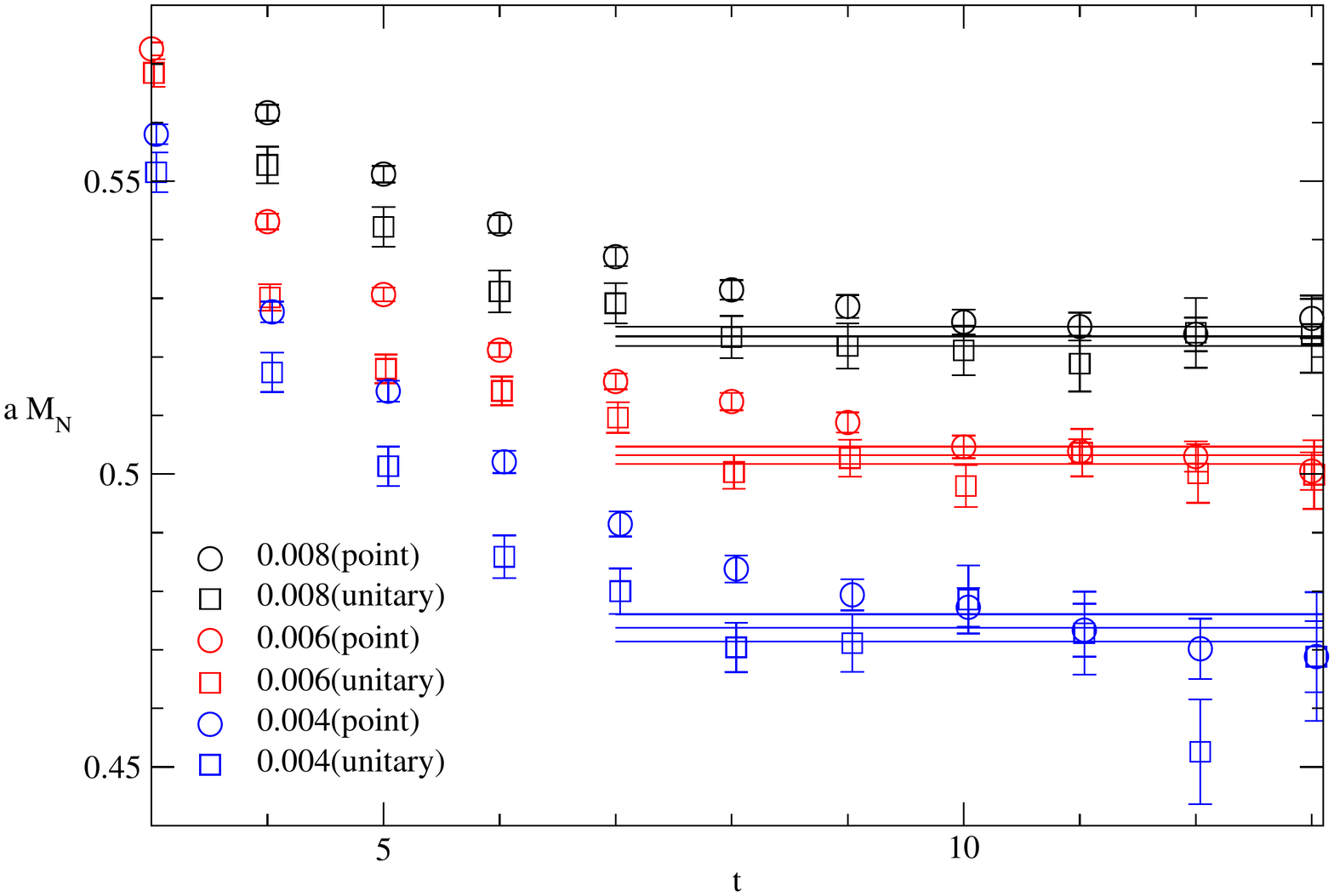}
\emin
\hspace{-0.05\textwidth}
\bmin
\includegraphics[angle=0,width=1.0\textwidth]{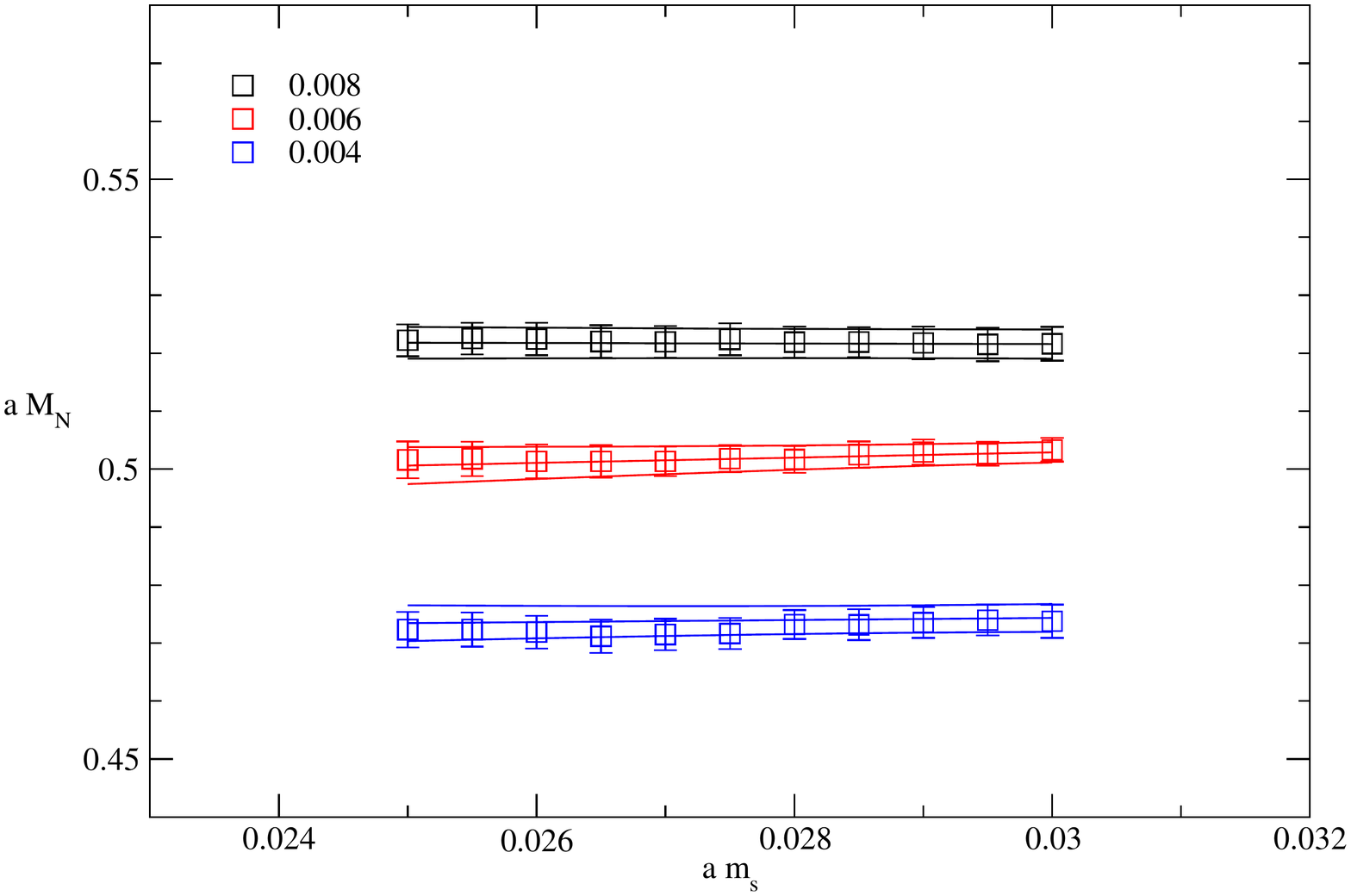}
\emin
\vspace{-0.05\textwidth}
\caption{ Nucleon effective mass and reweighted mass for DWF+I 0.08fm ensembles}
\label{fig:32}
\end{figure}

\begin{figure}[hbt]
\vspace{-0.05\textwidth}
\hspace{-0.05\textwidth}
\bmin
\includegraphics[angle=0,width=1.0\textwidth]{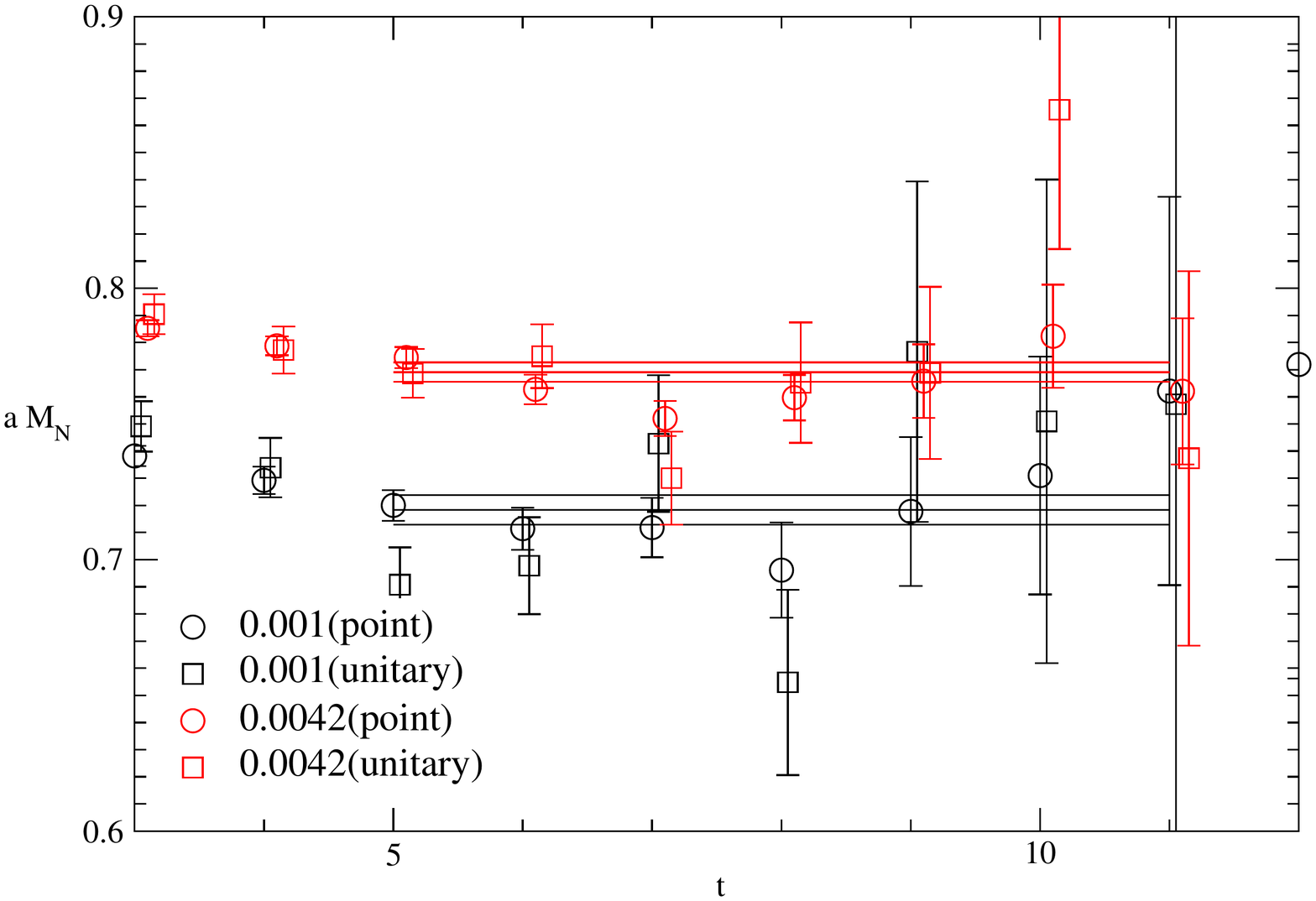}
\emin
\hspace{-0.05\textwidth}
\bmin
\includegraphics[angle=0,width=1.0\textwidth]{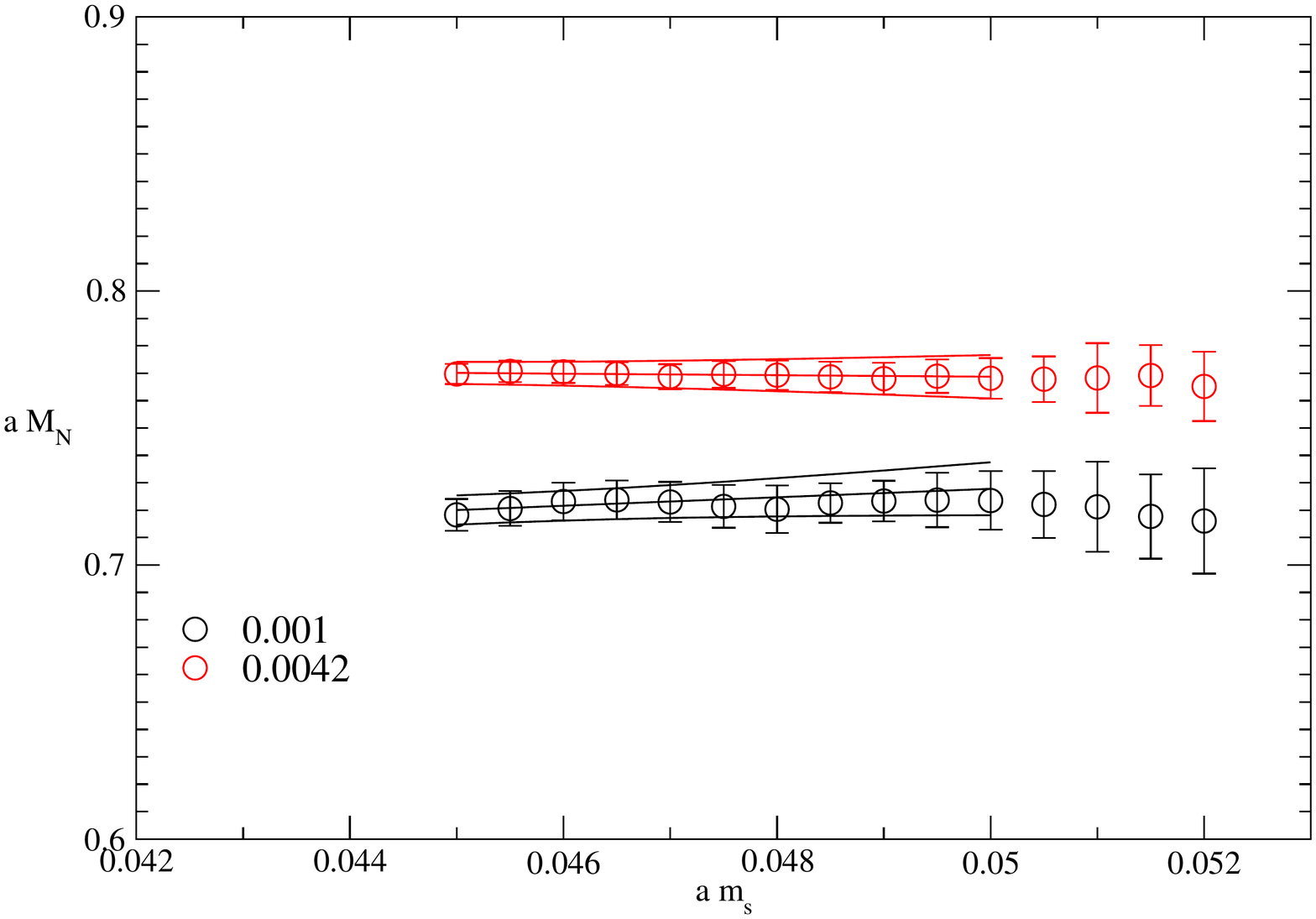}
\emin
\vspace{-0.05\textwidth}
\caption{ Nucleon effective mass and reweighted mass for DWF+ID ensembles}
\label{fig:DSDR}
\end{figure}
Figures~\ref{fig:24} to~\ref{fig:DSDR} shows the reweighted nucleon mass $M_N(m'_s)$, as well as the fitting range used in extracting $\NssN(m_l,a)$ for DWF+I and DWF+ID ensembles.
A preliminary extrapolation to the continuum limit and physical pion mass gives $\NssN=0.33(31)$ at 2Gev with $\overline{MS}$ while fitting without $a^2$ gives 0.09(16), which in turn gives
\be
\fTs = 0.035(33) (\mbox{with } a^2),\quad\ 0.009(17) (\mbox{wit out } a^2).
\ee

\begin{figure}[hbt]
\begin{center}
\includegraphics[angle=0,width=1.0\textwidth]{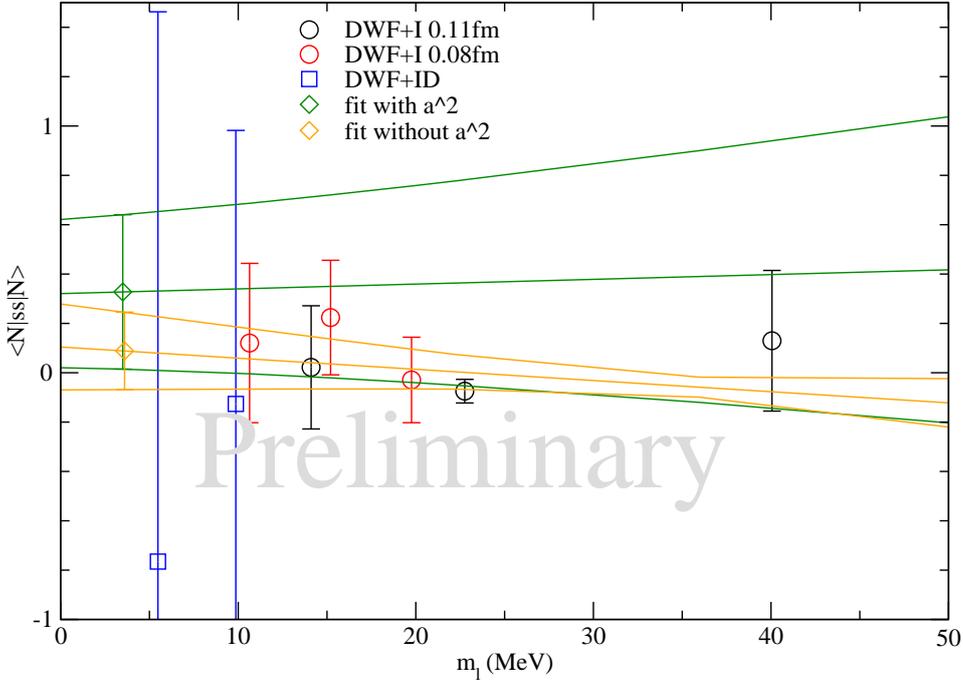}
\end{center}
\vspace{-0.1\textwidth}
\caption{Nucleon strange contents calculated via Feynman-Hellman theorem and mass reweighting, renormalized at 2Gev.}
\label{fig:NssN}
\end{figure}
\section{Discussion}
We reported on the mass and the strange quark content of the nucleon, measured on the (2+1)-flavor dynamical DWF ensembles generated by RBC/UKQCD collaborations. Having DWF+I ensembles with  2 different lattice spacings allows us to do a continuum extrapolation. 
Comparing the continuum value from fits with and without $a^2$ terms suggests the lattice spacing error could be larger than previously estimated, and could explain the apparent discrepancy between different lattice studies. 
While we did generate new propagators on DWF+I 0.11fm ensembles, 
this results presented in this paper were  obtained by mostly re-using propagators and reweighting 
factors generated for other studies \cite{ Aoki:2010dy, Arthur:2012yc, Syritsyn:2009mx}. 

We also analyzed DWF+ID ensembles generated with Dislocation Suppressing Determinant Ratio(DSDR), which currently has the lightest pions and largest volumes for dynamical DWF ensembles. The results with the currently available statistics are still too noisy and the results were not included in the continuum extrapolation of $\NssN$.

It appears quite possible that further optimization of sources for the measurement of nucleon mass as well as increased statistics can improve the signal for $\NssN$ significantly without the need to extend the ensemble.
While generating propagators with different sources and/or positions traditionally has required completely separate inversions and made the numerical cost prohibitively expensive, recent developments of various techniques such as EigCG~\cite{Stathopoulos:2007zi}. Low/All mode averaging(LMA/AMA)~\cite{Blum:2012uh} makes it possible to generate a large number of propagators per configurations at a numerical cost only a few times the cost for single propagator. Nucleon studies are particularly well suited to take advantage of this, as the correlation length for nucleons are much shorter than those for the light mesons. An exploratory study with AMA is under way.

\vspace{0.05\textwidth}
C.J was supported by the US DOE under contract DE-AC02-98CH10886.
The gauge configurations used in this study were generated at QCDOC machines at Brookhaven National Laboratory and Edinburgh Parallel Computing Center(EPCC), and IBM BG/P machines at Argonne Leadership Computing Facility(ALCF). The reweighting factors and propagators were generated at ALCF, RIKEN Integrated Cluster of Clusters(RICC) and NSF Teragrid/XSEDE facilities.

\providecommand{\href}[2]{#2}
\begin{spacing}{0.8}

\end{spacing}

\end{document}